\documentclass[10pt,conference]{IEEEtran}
\usepackage{graphicx} 

\newtheorem{definition}{Definition}[section]

\newtheorem{theorem}{Theorem}[section]

\begin{document}

\title{Near Perfect Decoding of LDPC Codes}

\author{\authorblockN{Xiaofei Huang}
\authorblockA{Coding Research\\
Foster City, California, USA \\
Email: huangxiaofei@ieee.org}
}
%

\maketitle

\begin{abstract}
Cooperative optimization is a new way 
	for finding global optima of complicated functions of many variables.
It has some important properties 
	not possessed by any conventional optimization methods.
It has been successfully applied in solving many large scale optimization problems 
	in image processing, computer vision, and computational chemistry.
This paper shows the application of this optimization principle
	in decoding LDPC codes, which is another hard combinatorial optimization problem. 
In our experiments, it significantly out-performed the sum-product algorithm, 
	the best known method for decoding LDPC codes.
Compared to the sum-product algorithm,
	our algorithm reduced the error rate further by three fold, 
	improved the speed by six times, 
	and lowered error floors dramatically in the decoding.
\end{abstract}

\section{Introduction}
Decoding plays a very important role in modern data communications.
The best known decoding algorithm for Turbo codes and LDPC codes
	is called the sum-product algorithm~\cite{Kschischang01}.
It is a message passing algorithm operating in a general graph.
To the surprise of many mathematicians, 
	we have little theoretical understanding of its principles despite of its effectiveness.
Although it has demonstrated satisfying performances in decoding and solving many other optimization problems,
	it may also give poor results or fail to converge.
	
This paper presents the application of cooperative optimization in decoding LDPC codes.
It is a new optimization principle completely unknown to the mathematics and engineering societies before.
Similar to the sum-product algorithm, 
	the cooperative algorithm also employs message passing operated in a general graph. 
Unlike the sum-product algorithm,
	its computational properties are better understood.
While many classic methods struggled with local minima, 
	our method always has a unique equilibrium and converges to it 
	with an exponential rate regardless of initial conditions.	
It can determine if the equilibrium is a global optimum or not.
In many important cases, it guarantees to find the global optima for difficult optimization problems
	when conventional methods often fail.

Theories for optimization have been studied for centuries. 
They caught special attention after the invention of computers 
	because of their importance in solving many practical problems with the use of computers.
Yet in the past many effective optimization methods are found not by applying the known optimization theories.
Instead they are empirical ones discovered with some threads of chances, 
	just like the sum-product algorithm for decoding Turbo codes and LDPC codes used in data communications.
This crucial realization demands us to discover new principles for optimization and build new theories for them.
We can always expect better results through deeper theoretical understanding beyond discovering empirical rules.
Hopefully, the application of a new optimization principle for decoding LDPC codes presented in this paper
	could support this point of view.

\section{The Cooperative Optimization Principle}

\subsection{Basic Ideas}

To solve a hard problem, 
   we follow the divide-and-conquer principle.
We first break up the problem into a number of sub-problems of
   manageable sizes and complexities.
Following that, we assign each sub-problem to an agent,
	and ask those agents to solve the sub-problems in a cooperative way.
The cooperation is achieved by asking each agent
   to compromise its solution with the solutions of others
   instead of solving the sub-problems independently.
We can make an analogy with team playing, 
   where the team members work together
   to achieve the best for the team, but not necessarily the best for each member.
In many cases, cooperation of this kind can dramatically improve the problem-solving capabilities
	of the agents as a team, even when each agent may have very limited power.
	
To be more specific, the cooperation is achieved in such a multi-agent system
	via two vital steps executed by each agent in an iterative way,
	1) solving its sub-problem by soft decision making, and 
	2) passing its soft decisions to its neighboring agents.
At the very beginning, 
	each agent makes soft decisions by solving its own sub-problem 
	and ranking the solutions in order of preferences measured by some values.
For an agent, the most preferable one is the best solution 
	to its sub-problem and the less preferable ones 
	are the solutions sub-optimal to its sub-problem.
Following that, each agent passes its soft decisions as messages to its neighboring agents.
After receiving its neighbor agents' soft decisions, 
	each agent goes back to the soft decision making step again.
At this time, instead of solving its sub-problem independently,
	it tries to solve its sub-problem by compromising its solutions with its neighboring agents'.
The best solution for one agent may not be the best one for another.
If there is any conflict among the agents,
	it is required for each agent to compromise its solutions with its neighbors to reach a consensus.
If a consensus in picking solutions is reached through compromising, 
	the system reports it as a solution for the original problem.
Otherwise, the system iterates until a consensus is reached among the agents
	or the iteration exceeds some cap.
	
The very core of cooperation is the soft decision making via solution compromising.
Paper~\cite{HuangBookCCO} formally describes the cooperative optimization in the language of game theory.
It has been shown in \cite{HuangBookCCO} that there are different cooperation schemes 
	yielding different computational behaviors of the system.
One of them leads the system to find the Nash equilibria.
Another ensures the system of a unique equilibrium.
With this scheme, the system always converges to the equilibrium
	with an exponential rate regardless of initial conditions.	
Theory also tells us that the equilibrium must be the global optimum 
	if it is a consensus solution.
Details about these together with the theoretical investigation of the cooperative optimization
	are provided in \cite{HuangBookCCO}.

\subsection{A Simple Example}	
Let the cost function (also referred to as energy function or objective function) to be minimized be $E(x_1, x_2, x_3)$, 
	which can be expressed as an aggregation
\begin{equation}
E(x_1, x_2, x_3) = f_{12}(x_1, x_2) + f_{23}(x_2, x_3) +f_{13}(x_1, x_3)
\label{cost_function}
\end{equation}
of three binary sub-functions, $f_{12}(x_1, x_2)$, $f_{23}(x_2, x_3)$, and $f_{13}(x_1, x_3)$.

\begin{figure}
\centering
\includegraphics[width=7.8cm]{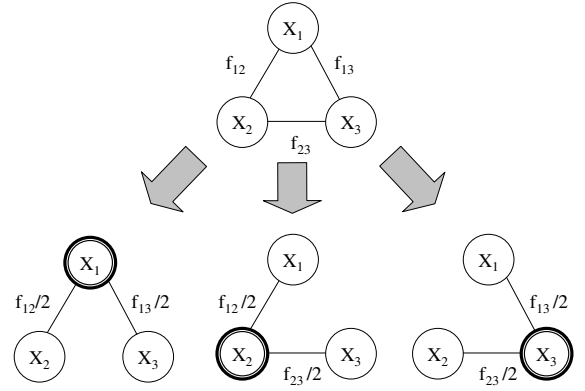}
\caption{The illustration of decomposing a graph with loop(s) into sub-graphs of tree-like structures.}
\label{fig_decomposition}
\end{figure}

To illustrate the decomposition of this problem into simple sub-problems,
	we map the cost function (\ref{cost_function}) into a graph (shown in the upper portion of Fig.~\ref{fig_decomposition}).
We can view each variable as a node in the graph and 
   each binary sub-function as a connection between two nodes.
This graph has one loop and we can decompose it into three sub-graphs of no loop shown in the lower portion of  Fig.~\ref{fig_decomposition},
	one for each variable (double circled).
Each sub-graph is associated with one cost function, $E_i, i = 1, 2, 3$.
For example, the sub-graph for variable $x_1$ has its cost function $E_1$ as
\[ E_1 (x_1, x_2, x_3) = (f_{12} (x_1, x_2) + f_{13}(x_1, x_2))/2 \ . \]
So are the cost functions of the sub-graphs for other two variables:
\[ E_2 (x_1, x_2, x_3) = (f_{23} (x_2, x_3) + f_{12}(x_1, x_2))/2 \ , \]
\[ E_3 (x_1, x_2, x_3) = (f_{13} (x_1, x_3) + f_{23}(x_2, x_3))/2 \ . \]
Obviously,
\[E= E_1 + E_2 + E_3 \ . \]
With such a decomposition, the original problem, $\min E$, becomes
	three sub-problems, $\min E_i$, $i = 1, 2, 3$.

For the $i$th sub-problem, 
	the preferences for picking values for variable $x_i$ are used as the soft decisions 
	for solving the sub-problem.
Those preferences are measured by some real values and are described as a function of $x_i$, 
	denoted as $c_i(x_i)$.
It is also called the assignment constraint for variable $x_i$.
The different function values, $c_i(x_i)$, 
	stand for the different preferences in picking values for variable $x_i$.
Because we are dealing with minimizing $E$, for the convenience of the mathematical manipulation,
	we choose to use smaller function values, $c_i(x_i)$s,  for more preferable variable values.
	
To introduce cooperation in solving the sub-problems,
	we iteratively update the assignment constraints (soft decisions in assigning variables) as
\begin{eqnarray}
c^{(k)}_1 (x_i) = \min_{x_2,x_3} \left(1 - \lambda_k \right) E_1  +
\lambda_k \sum_{j} w_{1j} c^{(k-1)}_j(x_j) \label{u1} \\ 
c^{(k)}_2 (x_i) = \min_{x_1,x_3} \left(1 - \lambda_k \right) E_2  +
\lambda_k \sum_{j} w_{2j} c^{(k-1)}_j(x_j)  \label{u2}\\
c^{(k)}_3 (x_i) = \min_{x_1,x_2} \left(1 - \lambda_k \right) E_3  +
\lambda_k \sum_{j} w_{3j} c^{(k-1)}_j(x_j) \label{u3}
\end{eqnarray}
where $k$ is the iteration step, 
   $w_{ij}$ are non-negative weight values satisfying $\sum_i w_{ij} = 1$.
It has been found~\cite{HuangBookCCO} that such a choice of $w_{ij}$ makes sure 
   the iterative update functions converge. 

Parameter $\lambda_k$ controls the level of the cooperation at step $k$
and is called the cooperation strength, satisfying $0 \le \lambda_k < 1.$   
A higher value for $\lambda_k$ will weigh 
   the solutions of the other sub-problems $c_j(x_j)$
   more than the one of the current sub-problem $E_i$.
In other words, the solution of each sub-problem
   will compromise more with the solutions of other sub-problems.
As a consequence, 
   a higher level of cooperation in the optimization is reached in this case.
   
The update functions, (\ref{u1}),(\ref{u2}), and (\ref{u3}), are a set of difference equations of 
   the assignment constraints $c_i(x_i)$.
Unlike conventional difference equations used by probabilistic relaxation algorithms~\cite{Rosenfeld76},
   and Hopfield Networks~\cite{Hopfield82},
   this set of difference equations always has one and only one equilibrium given $\lambda$ and $w_{ij}$.
Some important properties of this cooperative optimization 
	will be shown in the following subsections.

\subsection{Cooperative Optimization in a General Form}

Let $E(x_1,x_2, \ldots, x_n)$ be a multivariate cost function, or simply denoted as $E(x)$,
   where each variable $x_i$ has a finite domain $D_i$ of size $m_i$ ($m_i=|D_i|$).
We break the function into $n$ sub-cost functions $E_i$ ($i=1,2,\ldots,n$),
	one for each variable, 
   such that $E_i$ contains at least variable $x_i$, 
   the minimization of each cost function $E_i$ (the sub-problem)
   is computational manageable in complexity, and
\begin{equation}
E (x) = \sum^{n}_{i = 1} E_i (x).
\label{decomposition}
\end{equation}

The cooperative optimization is defined by the following set of difference equations:
\begin{equation}
c^{(k)}_i (x_i) = \min_{x_j \in  X_i \setminus{x_i}}\left( \left(1 - \lambda_k \right) E_i  +
\lambda_k \sum_{j} w_{ij} c^{(k-1)}_j(x_j)\right) \ . 
\label{cooperative_optimization}
\end{equation}

Intuitively, we might choose $w_{ij}$ 
   such that it is non-zero if $x_j$ is contained by $E_i$.
However, theory tells us that this is too restrictive.
To make the algorithm work,
   we need to choose 
   $(w_{ij})_{n \times n}$ to be a propagation matrix defined as follows:

\begin{definition}
A propagation matrix $W = (w_{ij})_{n \times n}$
   is a irreducible, nonnegative, real-valued square matrix and satisfies
\[ \sum^n_{i=1} w_{ij}=1, \quad \mbox{ for } 1 \le j \le n\ . \]
\label{definition_propagation_matrix}
\end{definition}

A matrix $W$ is called reducible if there exists
   a permutation matrix $P$ such that $PWP^T$
   has the block form
\[\left(
    \begin{array}{cc}
      A & B \\
      O & C 
    \end{array}
\right)\ . \]  

\begin{definition}
The system is called reaching a consensus solution
   if, for any $i$ and $j$ where $E_j$ contains $x_i$, 
   \[ \arg \min_{x_i} \tilde{E}_i = \arg \min_{x_i} \tilde{E}_j \ , \]
where $\tilde{E}_i$ is defined as
\[\tilde{E}_i = \left(1 - \lambda_k \right) E_i  + \lambda_k \sum_{j} w_{ij} c^{(k-1)}_j(x_j) \ . \]
\label{consensus}
\end{definition}

\begin{definition}
A solution to the difference equations~(\ref{cooperative_optimization})
   is called an equilibrium of the system.
Specifically, it is a set of values for all the assignment constraints (the soft decisions), 
	$(c_1(x_1), c_2(x_2), \ldots, c_n(x_n) )$,
   such that the difference equations are satisfied.
\end{definition}

To simplify the notations in the following discussions,
   let
\[ c^{(k)} = (c^{(k)}_1, c^{(k)}_2, \ldots,c^{(k)}_n). \]
Let $\tilde{x}^{(k)}_i = \arg \min_{x_i} c^{(k)}_i(x_i)$, 
   the favorable value for assigning variable $x_i$.
Let $\tilde{x}^{(k)} = (\tilde{x}^{(k)}_1, \tilde{x}^{(k)}_2, \ldots, \tilde{x}^{(k)}_n)$.
It is the candidate solution obtained by the cooperative algorithm at iteration $k$.
   
\subsection{Some Important Properties}

The theoretical understanding of the cooperative optimization 
	has been given in detail in \cite{HuangBookCCO}.
Here we list some important properties.

The following theorem shows that 
   $c^{(k)}_i(x_i)$ for $x_i \in D_i$
   have a direct relationship to the lower bound on the cost function $E(x)$.

\begin{theorem} 
Given any propagation matrix $W$
   and the general initial condition $c^{(0)}=0$ or $\lambda_{1}=0$,
   $\sum_i c^{(k)}_i(x_i)$ is a lower bound function on $E(x_1,\ldots, x_n)$,
   denoted as $E^{(k)}_{-}(x_1,\ldots, x_n)$.
That is 
\begin{equation}
\sum_i c^{(k)}_i(x_i) \le E(x_1,x_2,\ldots, x_n),~~~~~ \mbox{for any $k\ge 1$}\ .
\label{up_bound}
\end{equation}
In particular, let $E^{*(k)}_{-}=\sum c^{(k)}_i(\tilde{x_i})$, 
   then $E^{*(k)}_{-}$ is a lower bound on the optimal cost $E^{*}$, 
   that is $E^{*(k)}_{-} \le E^{*}$.
\label{theorem_1}
\end{theorem}

Here, subscript ``-'' in $E^{*(k)}_{-}$ indicates that
   it is a lower bound on $E^{*}$.

This theorem tells us that $\sum c^{(k)}_i(\tilde{x_i})$ provides
   a lower bound on the cost function $E$.
We will show in the next theorem that
   this lower bound is guaranteed to be improved 
   as the iteration proceeds.
\begin{theorem}
Given any propagation matrix $W$,
   a constant cooperation strength $\lambda$,
   and the general condition $c^{(0)}=0$, 
   $\{E^{*(k)}_{-}|k \ge 0\}$
   is a non-decreasing sequence with upper bound $E^{*}$.
\label{theorem_3}
\end{theorem}

If a consensus solution is found at some step or steps,
   then we can find out the closeness between 
   the consensus solution and the global optimum in cost.
If the algorithm converges to a consensus solution, 
   then it must be the global optimum also.
The following theorem makes these points clearer.

\begin{theorem}
Given any propagation matrix $W$,
   and the general initial condition $c^{(0)}=0$ or $\lambda_1=0$.
If a consensus solution $\tilde{x}$ is found
   at iteration step $k_1$ and remains the same from step $k_1$ to step $k_2$,
then the closeness between the cost of $\tilde{x}$,
    $E(\tilde{x})$, and the optimal cost, $E^{*}$, satisfies 
   the following two inequalities,
\begin{equation}
0 \le E(\tilde{x})- E^{*} \le 
   \left(\prod^{k_2}_{k=k_1} \lambda_k\right) \left(E(\tilde{x})-E^{*(k_1-1)}_{-}\right),
\end{equation}
\begin{equation}
0 \le E(\tilde{x})- E^{*} \le 
   \frac{\prod^{k_2}_{k=k_1} \lambda_k}{1-\prod^{k_2}_{k=k_1}\lambda_k} (E^{*}-E^{*(k_1-1)}_{-})\ ,
\end{equation}
where $(E^{*}-E^{*(k_1-1)}_{-})$ is the difference 
   between the optimal cost $E^{*}$ and
   the lower bound on the optimal cost, $E^{*(k_1-1)}_{-}$,
   obtained at step $k_1 - 1$.
When $k_2-k_1 \rightarrow \infty$ and $1 - \lambda_k \ge \epsilon > 0$ 
   for $k_1 \le k \le k_2$, $E(\tilde{x}) \rightarrow E^{*}$.
\label{theorem_2}
\end{theorem}  

The performance of the cooperative algorithm further depends on
   the dynamic behavior of the difference equations~(\ref{cooperative_optimization}).
Its convergence property
   is revealed in the following two theorems.
The first one shows that, 
   given any propagation matrix 
   and a constant cooperation strength,
   there does exist a solution
   to satisfy the difference equations (\ref{cooperative_optimization}).
The second part shows that the cooperative algorithm 
   converges exponentially to that solution.

\begin{theorem}
Given any symmetric propagation matrix $W$
   and a constant cooperation strength $\lambda$,
   then Difference Equations~(\ref{cooperative_optimization})
   have one and only one solution, denoted as $(c^{(\infty)}_i(x_i))$
   or simply $\mbox{\boldmath c}^{(\infty)}$.
\label{theorem_7}
\end{theorem}

\begin{theorem}
Given any symmetric propagation matrix $W$ and 
   a constant cooperation strength $\lambda$,
   the cooperative algorithm,
   with any choice of the initial condition $c^{(0)}$,
   converges to $c^{(\infty)}$ with an exponential convergence rate $\lambda$.
That is
\begin{equation}
\|c^{(k)}-c^{(\infty)}\|_{\infty} \le 
   \lambda^k \|c^{(0)}-c^{(\infty)}\|_{\infty}\ .
\end{equation}
\label{theorem_8}
\end{theorem}

This theorem is called the convergence theorem.
It indicates that
   our cooperative algorithm is stable and
   has a unique attractor, $c^{(\infty)}$.
Hence, the evolution of our cooperative algorithm is robust,
   insensitive to perturbations, and
   the final solution of the algorithm
   is independent of initial conditions.
In contrast, conventional algorithms
   based on iterative improvement (e.g. gradient descent)
   have many local attractors due to the local minima problem.
The evolution of these algorithms are sensitive 
   to perturbations,
   and the final solutions of these algorithms
   are dependent on initial conditions.
   
\section{Decoding LDPC via Cooperative Optimization}

\subsection{LDPC codes}
LDPC codes belong to a special class of linear block codes
	whose parity check matrix $H$ has a low density of ones.
LDPC codes were originally introduced by Gallager in his thesis~\cite{Gallager:LDPCC:thesis}.
After the discovery of turbo codes in 1993 by Berrou et al.~\cite{Berrou93},
	LDPC codes were rediscovered by Mackay and Neal~\cite{MacKay:GCBOVSM} in 1995.
Both classes have excellent performances in terms of error correction close to the Shannon limit.

The parity check matrix $H$ is a binary matrix with elements in $\{0,1\}$.
It is sparse with a few non-zero elements.
Let the code word length be $n$ and the input data be 
\[ x= ( x_1, x_2, \ldots, x_n ) \ , \]
then $H$ is a $n \times k$ matrix, where $k$ is the number of rows. 
Each row of $H$, denoted as $H_j$, introduces one parity check constraint on $x$,
\[ H_j x^T = 0~mod~2 \ . \]
Since $H$ has $k$ rows, 
	there are $k$ constraints on $x$.
That is,
\[ H x^T = 0~mod~2 \ . \]

\subsection{Maximum-Likelihood Decoding}
To minimize the probability of decoding error, the optimal decoder for a channel code
	finds an input $x$ that has the maximum posterior probability $P\{x | y\}$ given an output $y$.
Usually, we assume a uniform prior distribution on $x$. 
In this case, the maximum posterior criteria reduces to the maximum likelihood,
	i.e., finding an input $x$ which makes the likelihood distribution $P\{y | x\}$ a maximum.

For a discrete memoryless additive Guassian channel and a binary modulation,
	the output data bit at position $i$, $y_i$, 
		can be modeled as the following random variable:
\[ y_i = (2x_i - 1) + \xi_i \ , \]		
where $x_i \in \{0, 1\}$ and $\xi_i$ is a additive noise of the Gaussian distribution with variance $\sigma^2$.

Let 
\[ a_i = Log \frac{P\{y_i/x_i = 1\}}{P\{y_i / x_i = 0\}} \ , \]
where $P\{y_i/x_i\}, x_i = 0, 1$, is the conditional distribution of output data bit $y_i$ given the input data bit $x_i$.
In this case, the maximum likelihood decoding becomes
\begin{equation}
\max_{x_1, x_2, \ldots, x_n} \sum_i a_i (2x_i - 1), \quad \mbox{ s.t. $Hx = 0$ mod $2$} \ . 
\label{original_problem}
\end{equation}

This is a constrained maximization problem. 
Without loss of generality, we can transform it into an unconstrained minimization problem in a more general form.
To do that, we introduce unary constraints on variables $x_i$,
\[ f_i(x_i) = \left\{ \begin{array}{ll}
                      -2a_i & \mbox{ if $x_i = 1$} \\
                       2a_i & \mbox{ if $x_i = 0$} 
                     \end{array}
             \right. \]
and convert each parity check constraint to a $m$-ary constraint on the variables of the constraint.
Let the $j$th parity check constraint, $H_j x^T = 0~mod~2$, 
	define on a subset of variables, denoted as $X_j$ of size $|X_j| = m_j$.
Then the $m$-ary constraint on $X_j$ is defined as 
\[ f_{X_j}(X_j) = \left\{ \begin{array}{ll}
                      0 & \mbox{if $H_j x^T = 0~mod~2$} \\
                      \infty & \mbox{ otherwise} 
                     \end{array}
             \right. \]
Using those definitions, 
	(\ref{original_problem}) becomes 
\begin{equation}
\min_{x_1, x_2, \ldots, x_n} \sum_i f_i(x_i) + \sum_j f_{X_j} (X_j) \ . 
\label{optimization_problem}
\end{equation}
In general, the above problem is called the constraint-based optimization,
	which is NP-hard.
It is a core problem in mathematical logic and computing theory. 
In practice, it is fundamental in solving many problems in machine vision, 
	image processing, computational chemistry, integrated circuit design,
	computer network design, artificial intelligence and more.
	
\subsection{Decomposing into Sub-problems}

The Tanner graph is used to help us understand the decomposition.
A Tanner graph for a LDPC code is a bipartite graph with variable nodes on one side
	and constraint nodes (parity check nodes) on the other side. 
	Edges in the graph connect constraint nodes to variable nodes. 
A constraint node connects to those variable nodes that participate in its parity check.
A variable node connects to those constraint nodes that use the variable in the parity checks.

The Tanner graph can be decomposed into $n$ tree-like sub-graphs, 
	one for each variable.
Those sub-graphs can have overlaps.
Because their tree-like structures, 
	we can find the exact solutions for the sub-problems associated with those sub-graphs.
There are many ways of decomposing a graph
	which lead to different performances
	of the cooperative algorithm.
A simple, straightforward way of decomposition is
	to have the sub-graph of each variable node consisting of
	all the constraint nodes linked to the variable node, together with their connections,
	all the variable nodes linked to those constraint nodes, together with their connections, 
	and the variable node itself.

\section{Experimental Results}
We developed high performance US patent pending methods and apparatus 
	for decoding Turbo codes and LDPC codes using the cooperative algorithm.
Usually, short LDPC codes have high commercial values because the decoding time is also short.
It also has been found that irregular LDPC codes have better performances than regular ones~\cite{Richardson:DOCAILDPCC}.

\begin{figure}
\centering
\includegraphics[width=8.6cm]{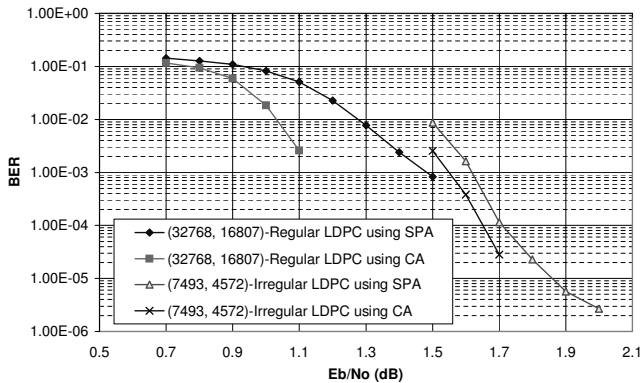}
\caption{Decoding an LDPC codes using SPA (the sum-product algorithm) and CA (the cooperative algorithm).}
\label{DecodingLDPC}
\end{figure}

A candidate code for China HDTV (proposed by the author using a new way of code construction called quantum coding) 
	is a $(7493, 4572)$-irregular LDPC code
	of a data rate $0.61$.
China decides to use LDPC codes instead of Turbo codes for channel coding 
	because of their higher coding gains and lower complexity in decoding. 
Fig~\ref{DecodingLDPC} shows the performances of the cooperative algorithm and 
	the sum-product algorithm in decoding the LDPC code using $10,000$ code words
	and AWGN (additive white Gaussian Noise) channel.

The maximum number of iterations for the sum-product algorithm is 30.
It was found that there is not much improvement in the decoding quality after 30 iterations.
An error floor was observed at BERs below $10^{-4}$ using the sum-product algorithm.
The acceptable error rate is below $10^{-9}$ for China HDTV.
The sum-product algorithm can not achieve that even after the $Eb/No$ is higher than $2.0~dB$

The error floor trouble was completely removed by the cooperative algorithm.
The error rate drops to zero after the $Eb/No$ is higher than $1.7~dB$.
At the "water fall" region, 
	the cooperative algorithm has reduced the decoding error rates further
	by more than three fold.
For the cooperative algorithm, 
	the maximum iteration number is 120 mainly because of much less complexity of its computation.
Even with that number, it was still more than six times faster than the sum-product algorithm.
	
In the second example, we use a regular LDPC code to demonstrate that 
	the cooperative algorithm has much less dependence on the code structure than the sum-product algorithm.
The code is the Turbo-like production of a simple parity check code $(8, 7)$
\[ D_1~D_2~D_3~D_4~D_5~D_6~D_7~P_8  \]
The configuration of the code is a $5$ dimensional cubic $(8, 7)^n$. 
The block size is $32768$, the data size is $16807$, and the data rate is $0.513$.
LPDC codes of this kind are simplest in structure and the most easy to encode (but not necessarily the best code distances).
The sum-product algorithm has terrible performance in decoding this kind of codes
	due to the high regularity of the code structure.
Fig~\ref{DecodingLDPC} shows the performances of both algorithms using $100$ code words and the AWGN channel.
The cooperative algorithm was much better than the sum-product algorithm in this case.
The success of the cooperative algorithm in decoding this type of LPDC codes
	implies that we can have greater flexibility at constructing high performance codes
	without worrying too much about the limitations of decoding algorithms.

\section{Conclusion}
We have presented the application of a new optimization technique 
	called cooperative optimization for decoding LDPC codes. 
Like the well-known sum-product algorithm, 
	the cooperative algorithm is also based on iterative message passing
   operating in the Tanner graph.
Although similar in operations, 
	they are derived from different principles. 

The sum-product algorithm is a generalization of the belief propagation algorithm~\cite{Pearl88} used in AI.
It can find exact solutions when the graph it operates on has no cycles.
With cycles, we still lack a theoretical understanding of the behavior of the algorithm.
Unlike many conventional methods, 
	cooperative optimization has a solid theoretical foundation on many computational properties.
In our experiments, it significantly outperformed the sum-product algorithm
	both in efficiency and accuracy for decoding different LDPC codes.
The new cooperative decoding algorithm can be extended further
	from the min-sum semiring to other semirings similar to those done for the sum-product algorithm~\cite{Kschischang01} 
	and general distributive law~\cite{Aji00}.

\end{document}